\documentclass[sigconf]{acmart}

\usepackage{multirow}

\AtBeginDocument{%
  \providecommand\BibTeX{{%
    \normalfont B\kern-0.5em{\scshape i\kern-0.25em b}\kern-0.8em\TeX}}}

\setcopyright{acmcopyright}

\copyrightyear{2023}
\acmYear{2023}
\setcopyright{acmlicensed}\acmConference[ISCA '23]{Proceedings of the 50th Annual International Symposium on Computer Architecture}{June 17--21, 2023}{Orlando, FL, USA}
\acmBooktitle{Proceedings of the 50th Annual International Symposium on Computer Architecture (ISCA '23), June 17--21, 2023, Orlando, FL, USA}
\acmPrice{15.00}
\acmDOI{10.1145/3579371.3589045}
\acmISBN{979-8-4007-0095-8/23/06}

\begin{document}

\title[HAAC: A Hardware-Software Co-Design to Accelerate Garbled Circuits]{HAAC: A Hardware-Software Co-Design \\to Accelerate Garbled Circuits}


\author{Jianqiao Mo}
\email{jm8782@nyu.edu}
\orcid{0000-0001-9533-8183}
\affiliation{%
  \institution{New York University}
  \city{New York}
  \state{New York}
  \country{USA}
}

\author{Jayanth Gopinath}
\email{jg6476@nyu.edu}
\orcid{0009-0006-8137-1250}
\affiliation{%
  \institution{New York University}
  \city{New York}
  \state{New York}
  \country{USA}
}

\author{Brandon Reagen}
\email{bjr5@nyu.edu}
\affiliation{%
  \institution{New York University}
  \city{New York}
  \state{New York}
  \country{USA}
}

\renewcommand{\shortauthors}{Jianqiao Mo, Jayanth Gopinath and Brandon Reagen}

\newcommand{\fixme}[1]{\textcolor{red}{#1}}
\newcommand{\hl}[1]{\textcolor{blue}{#1}}

\begin{abstract}
Privacy and security have rapidly emerged as priorities in system design. 
One powerful solution for providing both is
privacy-preserving computation, where functions are computed
directly on encrypted data and control can be provided 
over how data is used.
Garbled circuits (GCs) are a PPC technology that provide both confidential computing and control over how data is used.
The challenge is that they incur significant performance overheads compared to plaintext. 
This paper proposes a novel garbled circuits accelerator and compiler, named HAAC,
to mitigate performance overheads and make privacy-preserving computation more practical. 
HAAC is a hardware-software co-design.
GCs are exemplars of co-design as programs are completely known at compile time,
i.e., all dependence, memory accesses, and control flow are fixed.
The design philosophy of HAAC is to keep hardware simple and efficient,
maximizing area devoted to our proposed custom execution units and other circuits essential for high performance (e.g., on-chip storage).
The compiler can leverage its program understanding to realize hardware's performance potential by generating effective instruction schedules, data layouts, and orchestrating off-chip events.
In taking this approach we can achieve ASIC performance/efficiency without sacrificing generality.
Insights of our approach include how co-design enables expressing 
arbitrary GCs programs as streams, which simplifies hardware and enables complete memory-compute decoupling,
and the development of a scratchpad that captures data reuse by tracking program execution, eliminating the need for costly hardware managed caches and tagging logic.
We evaluate HAAC with VIP-Bench and achieve an average speedup of 589$\times$ with DDR4 (2,627$\times$ with HBM2) in 4.3mm$^2$ of area.
\end{abstract}




\maketitle

\section{Introduction}
\label{sec:introduction}

Privacy and security continue to increase in importance and demand new techniques to provide strong data protection guarantees.
This has given rise to a new paradigm of computing.
Privacy-preserving computation (PPC) can provide users two major advantages: confidentiality and control.
Confidential computing enables computation on encrypted data, guaranteeing that service providers cannot view users' sensitive, personal data while still providing them access to high-quality services.
Some techniques (namely secure multi-party computation) further allow users to control how their data is used, dictating which functions their data is computed with.
While promising, the ubiquitous deployment of all cryptographically strong PPC techniques are limited by high computational overheads, which today are too high for most applications.
Novel hardware solutions are needed to mitigate overheads and usher in a new era of private computing.

A variety of PPC techniques exist, this paper focuses on garbled circuits (GCs).
Each has its strengths and weaknesses, 
and we present them in detail in Section~\ref{sec:background}.
The future likely contains a mixture of techniques,
as their strengths can be combined to overcome limitations~\cite{ML_classificat_encrypted, MP-SPDZ}.
The intent of this paper is not to argue whether GCs or, for example,
homomorphic encryption is superior
but rather to show how the performance overheads of GCs can largely be overcome with hardware acceleration, 
bringing their strengths within reach.

GCs provide strong confidentiality guarantees and controls over how data is used.
A salient feature of GCs is support for arbitrary computation.
GCs programs constitute (secure) Boolean logic, 
implying any function can be implemented, including conditionals and floating point
(many alternative PPC techniques restrict functional support).
A notable, and motivational, application has been non-linear layers, e.g., ReLU, in private neural inference (PI)~\cite{gazelle, 244032Delphi, Chameleon, SecureML}.
Prior work has shown GCs are the primary bottleneck for PI in hybrid protocols,
which combine multiple PPCs to uphold high accuracy~\cite{karthiknew, 244032Delphi, minionn, ghodsi2021cryptonas}.
Recent work has even identified GCs acceleration as a key enabler of faster PI~\cite{karthiknew}.
The overheads accelerators must overcome are high: 
across eight VIP-Bench~\cite{vipbench} benchmarks
GCs are an average of 198,000$\times$ slower than plaintext.

Multiple factors contribute to GC's high overhead.
First, executing each GCs gate entails a significant amount of computation.
Note that GCs gates are cryptographic functions, distinct from a plaintext gate.
E.g., a single Boolean AND gate can involve four AES calls, two key expansions (similar to AES), and a variety of 128-bit logic operations.
Second, processing a function requires executing a large number of gates, as it must be expressed as Boolean logic.
For example, executing a private Bubble Sort from VIP-Bench requires processing over 12 million gates.
Third, GCs are data intensive~\cite{MAGE}.
Each plaintext gate's inputs and output are represented as a 128-bit ciphertext, 
and each (AND) gate involves a unique, 32 Byte, cryptographic constant for processing.

GCs have two fortuitous properties that enable effective hardware acceleration.
First, the core computations, though complex, are highly amenable to hardware implementation.
We show that by designing custom-logic hardware  leveraging parallelism within a gate performance can be significantly improved.
Second, the execution of a GCs program is entirely determined at compile time, 
providing software with a complete understanding of its behavior (i.e., a data oblivious program~\cite{vipbench}).
This presents a prime opportunity for hardware-software co-design.
We can develop programmable hardware (i.e., ISA support) for executing any GCs program with high performance and efficiency by relying on the compiler to organize \emph{all} data movement and instruction scheduling.
In doing so, gate computations can be parallelized across gate processing hardware while
data is streamed on-/off-chip to mask movement latency.
Reminiscent of VLIW, this eliminates costly hardware to extract performance from a program.
While the proposed hardware may be seen as simple,
this is intentional as it results in more area being devoted to the actual computation.
Alternative approaches are possible but tend to be overly restrictive or unnecessary.
Fixed-logic ASICs limit the arbitrary functional support GCs provide;
Systolic arrays and vectors constrain how data is laid out and computation ordered, 
which can restrict hardware's ability to process all programs well;
Dataflow is wasteful, as the compiler can handle scheduling and avoid allocating costly structures.

This paper presents HAAC, a novel co-design approach for accelerating GCs leveraging insights from their computational properties.
It includes a compiler, ISA, and hardware accelerator that combine to significantly improve GCs performance and efficiency.
HAAC accelerates individual gate execution with the gate engine (GE).
While GEs provide high performance potential, 
the challenge is exploiting gate parallelism and orchestrating data (operands, constants, instructions)
effectively while keeping hardware efficient.
A key insight is that the compiler, with hardware support, can express GCs as multiple streams.
First, the compiler can leverage instruction-level parallelism to improve intra/inter-GE parallel processing.
Knowing the precise order of events, instructions and constants can be streamed to each GE using queues.
Gate inputs and outputs, called \emph{wires}, 
are more difficult as they do not follow a linear pattern.
Streaming all wires on/off-chip is wasteful as they exhibit reuse.
Instead, we propose the sliding wire window (SWW) and wire renaming.
The SWW is a scratchpad memory that stores a contiguous address range of wires, 
and the range increases as the program executes.
Renaming is a complementary compiler pass that sequentializes gate output wire addresses according to program order.
The SWW and renaming combine to filter off-chip accesses, as recently generated wires are often soon reused in the circuit.
Thus, the SWW provides the performance benefits of a cache and efficiency of a scratchpad.
Most wire accesses are filtered, but misses, or Out-of-Range (OoR) accesses, still occur as capacity is limited.
OoR accesses cause significant performance degradation in HAAC's deep, in-order pipelines.
Our second wire optimization is to stream in OoR wires.
The insight is that the compiler knows when and which wires will be OoR
and can push them on-chip to an OoR wire queue.
An important implication of the optimizations is the enabling of complete decoupling of gate execution and off-chip accesses, allowing for total overlap.

This paper makes the following contributions:
\begin{enumerate}

    \item A novel hardware design tailored to GCs, including gate engines (GEs) to accelerate gates, queues (instruction, table, and OoR wire), and scratchpad (SWW).
    The hardware is programmable and supports an ISA.
    
    \item An optimizing compiler for parallel instruction scheduling (re-ordering), effective data layout and memory accesses (re-naming), and removing unnecessary off-chip communication (eliminating spent wires).
    
    \item A thorough evaluation with VIP-Bench\cite{vipbench}.
    With 16 GEs, a 2MB SWW, and DDR4, HAAC provides an average speedup of  
    589$\times$ (2,627$\times$ with HBM2) in 4.3mm$^2$.

\end{enumerate}

\begin{figure}[t]
\centerline{\includegraphics[width=\columnwidth]{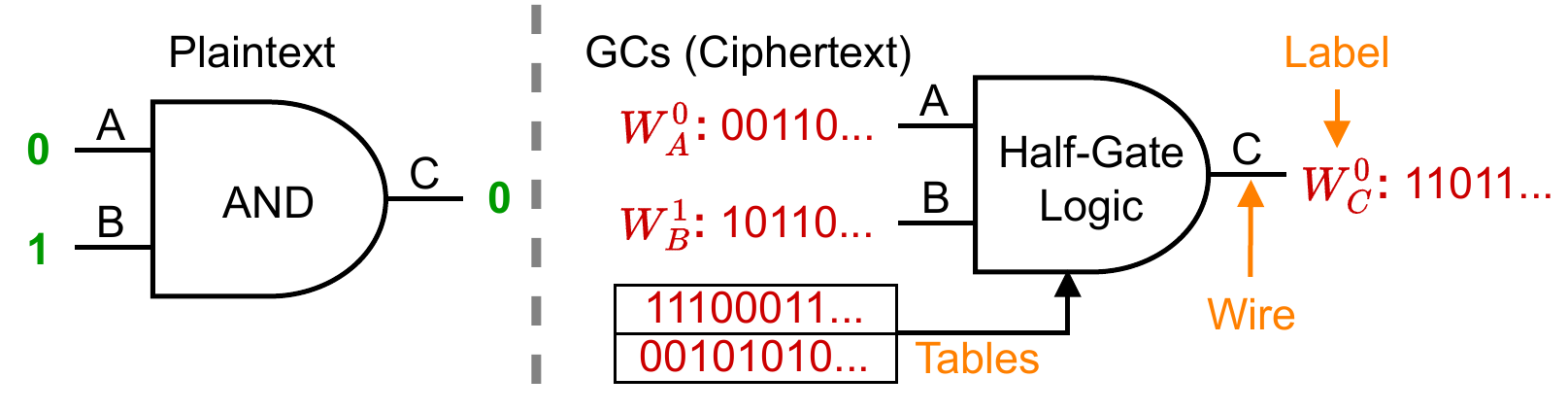}}
\vspace{-.6em}
\caption{
Wires (operands) are gate (operators) inputs and output. 
The values of wires are called labels and tables are constants used to evaluate gates.
}
\label{fig:gc_cartoon}
\vspace{-1em}
\end{figure}

\section{Background}
\label{sec:background}

This section provides a primer on GCs and compares PPC techniques.
The intent is to provide the reader with enough information to understand the contributions of the paper.
For a complete review, we refer those interested to related material~\cite{yao1986generate,10.1007/978-3-662-46803-6_8_HalfGate,inproceedingsFreeXOR,OT,emp-toolkit}.

\subsection{A Primer on GCs}

\textbf{Protocol:}
Garbled circuits are a type of secure two party computation with two phases: garbling and evaluation.
It allows two parties, Alice (Garbler) and Bob (Evaluator),
to jointly compute $y=f(a,b)$ on secret inputs: \textit{a} from Alice and \textit{b} from Bob. 
GCs support secure Boolean logic where operators are gates (typically AND and XOR), operands are called wires (i.e., gate inputs and outputs), and the encrypted values assigned to wires are called labels.
During garbling, one party (Alice, the Garbler) generates 
labels for all possible inputs.
Then, Bob obtains his labels corresponding to the values of \textit{b} from Alice without her learning anything about \textit{b} via oblivious transfer~\cite{OT}.
Alice also generates encrypted truth tables (constants) for each Boolean gate in the function $f$ and sends them to Bob.
As functions are known before inputs, garbling (label and table generation) can be done offline.
During the evaluation phase, Bob (Evaluator) takes the encrypted wire labels and tables as input to compute the function $f$, represented as a circuit of Boolean gates, securely.
Figure~\ref{fig:gc_cartoon} shows an example of a GCs gate evaluation and depicts key terms.

\textit{Garbling:}
GCs work by encrypting truth table representations of gates.
The first step is to convert a program $f$ into a Boolean netlist.
For each input wire $i\in(A, B)$, the Garbler will generate two 128-bit random labels, $W_i^{0}$ and $W_i^{1}$,
to represent the values 0 and 1, respectively.
Next, each gate in the netlist is encrypted as a garbled table.
The garbled table is generated by encrypting each output of the truth table twice using the labels corresponding to each row's inputs as keys.

\begin{figure}[t]
\centerline{\includegraphics[width=0.91\columnwidth]{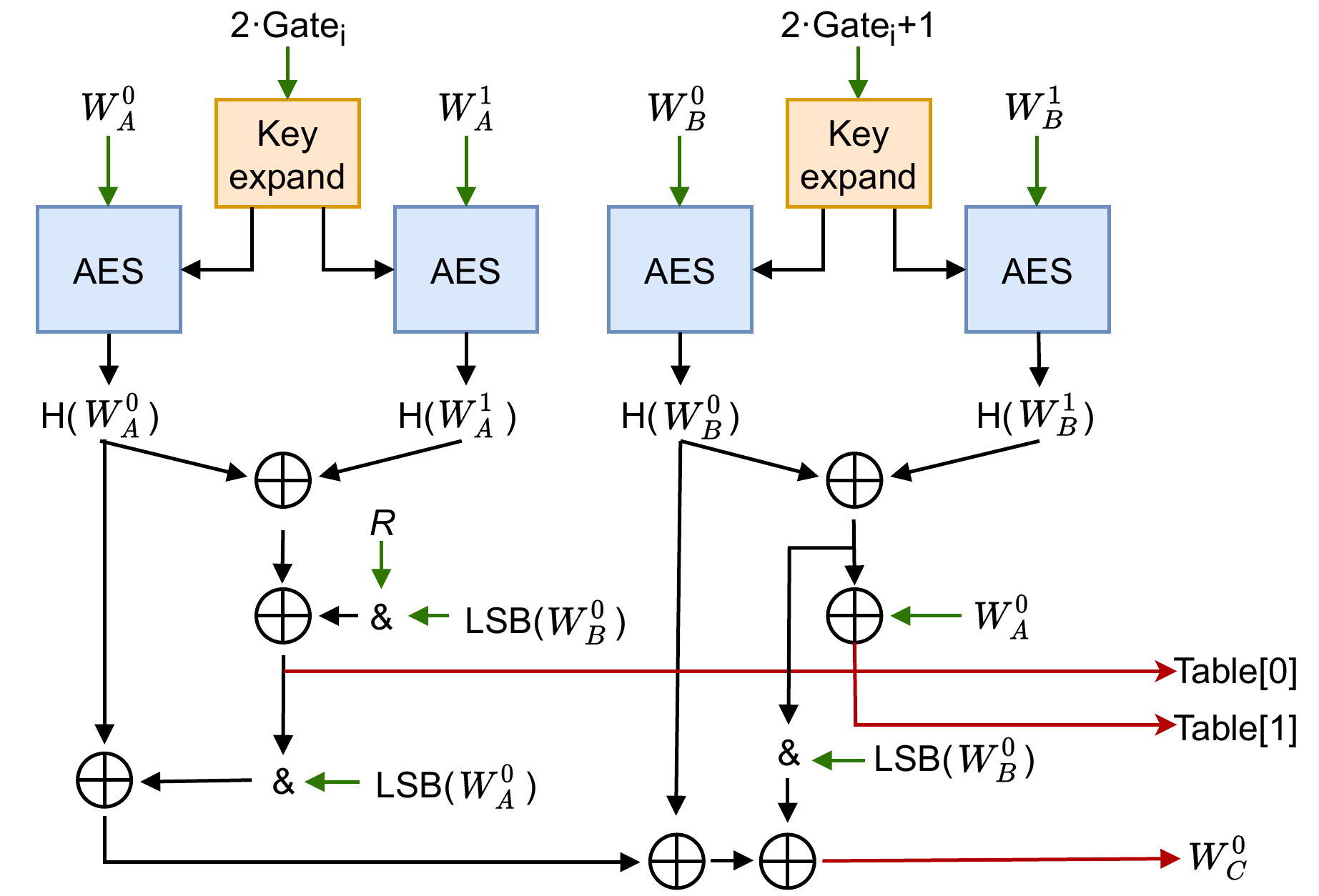}}
\caption{The Garbler's Half-Gate computation for AND.}
\label{fig:halfgate_gar}
\vspace{-1.5em}
\end{figure}

\textit{Evaluation:}
The evaluation phase uses the garbled tables to evaluate secure inputs.
To do this, plaintext data is converted to binary, and each input bit is mapped to its corresponding wire.
Then, wire labels $W_i$ are selected depending on the input values.
The Evaluator uses the selected labels as keys to decrypt each row of the table.
Recall each table row is encrypted twice.
With a unique label for each wire, and two wires per gate,
only one of the four rows will decrypt correctly, producing the desired gate output.
Once the Evaluator finishes processing all gates
the outputs are shared with the Garbler.

\textbf{GCs Gate:}
High-performance GCs constructions implement AND and XOR gates.
XOR is implemented using FreeXOR~\cite{inproceedingsFreeXOR} and 
AND using Half-Gate~\cite{10.1007/978-3-662-46803-6_8_HalfGate}.
We review these here as HAAC's gate engines implement them.

\textit{FreeXOR:}
FreeXOR enables the computation of XOR gates using only wire labels~\cite{inproceedingsFreeXOR}.
These gates are free in that they do not require garbled tables nor expensive AES computations to evaluate.
In FreeXOR, the Garbler generates a random
128-bit value $R$,
which is known only to herself. 
For each wire $i$, she generates the random label $W_i^{0}$ to represent logical 0
and sets $W_i^{1}= W_i^{0}\oplus R$ for logical 1.
The Garbler also sets output $W_C^{0}= W_A^{0}\oplus W_B^{0}$. 
Following this convention, an output wire label of C 
can be computed from input wires A and B as $W_C= W_A\oplus W_B$, 
without using a table.

\textit{Half-Gate:}
The Half-Gate is an efficient implementation of a GCs AND that halves the number of garbled table rows.
It has been proven as the optimal way of processing AND's garbled tables~\cite{10.1007/978-3-662-46803-6_8_HalfGate}.
We present the algorithm here as the Half-Gate is the primary functional unit in HAAC,
and HAAC stands for the \underline{HA}lf-Gate \underline{AC}celerator.
Figure~\ref{fig:halfgate_gar} illustrates the Garbler Half-Gate algorithm.
The Evaluator's process is similar, using half the number of AES calls.

The Garbler takes all labels of each wire as input to generate tables and output labels for following gates.
Each label is first run through a cryptographic hash function processed with AES, e.g., $H(W_{A}^{0})$.
Unlike the original GCs construction where labels are used as keys (as described above),
the Half-Gate uses the gate index as the key to construct the AES hash.
An important step here is key expansion, which is an AES like computation that expands the key, i.e., the gate index, to 176 Byte for AES operations.
The remaining logic at the bottom of the figure uses the hash output, labels, and $R$
to generate the tables and output wire label
($W_C^{0}$, and $W_C^{1} = W_C^{0}\oplus R$).
Details of the working and logic underlying these computations are beyond the scope of this review,
and complete details can be found in prior work~\cite{10.1007/978-3-030-56880-1_28}.

\textit{Fixed-key versus Re-keying:}
Fixed-key is an approach to reuse a key across multiple AES hashes,
reducing the key expansions per gate~\cite{fixed-key}.
This approach is used in prior work~\cite{8735500,Hussain2018MAXeleratorFA,Songhori2016GarbledCPUAM}.
However, recent work has shown using a fixed key reduces security~\cite{10.1145/2810103.2813619, 10.1007/978-3-030-56880-1_28}.
To maintain high security, HAAC uses \emph{re-keying} rather than fixed-key, 
processing full key expansions at extra computational cost.
Figure~\ref{fig:halfgate_gar} shows how secure AES hash functions are implemented using two distinct keys (instead of one in fixed-key).
We benchmark the difference and find re-keying increases the Half-Gate cost by 27.5\%.

\begin{table}[t]
\caption{
Comparison of PPCs considering:
confidentiality (Conf), 
data control (Cntrl), 
whether arbitrary compute is supported (Arb),
how security is achieved (Sec), 
level of performance overhead (Overhead), 
number of parties involved in the computation (Parties),
and whether the technology can execute without another PPC (Alone). }
\resizebox{\columnwidth}{!}
{
\setlength{\tabcolsep}{1.5mm}{
\begin{tabular}{cccccccc}
\hline
    & Conf & Cntrl & Arb & Sec   & Overhead  & Parties & Alone \\ \hline
HE  & Yes  & No    & No  & Noise & Very High & 1       & Yes   \\
TFHE & Yes  & No    & Yes & Noise & Ext. High & 1       & Yes   \\
SS  & Yes  & Yes   & No  & I.T.  & Moderate  & 2(+)    & No    \\
GCs  & Yes  & Yes   & Yes & AES   & Very High & 2       & Yes   \\ \hline
\end{tabular}
}
}
\vspace{-1em}
\label{tab:tradeoffs}
\end{table}

\subsection{Techniques for PPC}

We classify cryptographic privacy-preserving techniques into three categories: 
homomorphic encryption (HE), secret sharing (SS), and garbled circuits (GCs).
Each technique has strengths and weaknesses.
We overview them here and note that a common drawback is computational overhead.
A summary of tradeoffs can be found in Table~\ref{tab:tradeoffs}.

\textbf{Homomorphic encryption} works like standard encryption with the added benefit that functions can be computed directly on encrypted data for end-to-end confidentiality.
HE fits the mold of today's client-cloud service model,
requiring one party, typically the cloud, to process the computation.
It provides confidentiality but not control over how data is used.
Additionally, HE encryption is based on noise, which grows as computations are performed, and if it gets too large decryption fails.
This complicates programming as the user must manage noise growth.
Integer (including fixed point) HE schemes~\cite{BFV, BGV, ckks} only provide functional support for addition and multiplication, limiting what can be computed.
Boolean schemes exist (e.g., TFHE~\cite{tfhe}) that, like GCs, can compute arbitrary functions.
However, these incur extremely (Ext.) high performance overheads.
For example, a \emph{single} Boolean gate can take 75-600 milliseconds to process~\cite{TFHE-boots, TFHE-encoding}.
Integer HE slowdown is typically on the order of 5-6 orders of magnitude and
most systems research has been focused here~\cite{reagen2020cheetah,feldmann2021f1, bts, hpcafpga, riazi2020heax, craterlake}.

\begin{figure*}[t]
\centerline{\includegraphics[width=.95\textwidth]{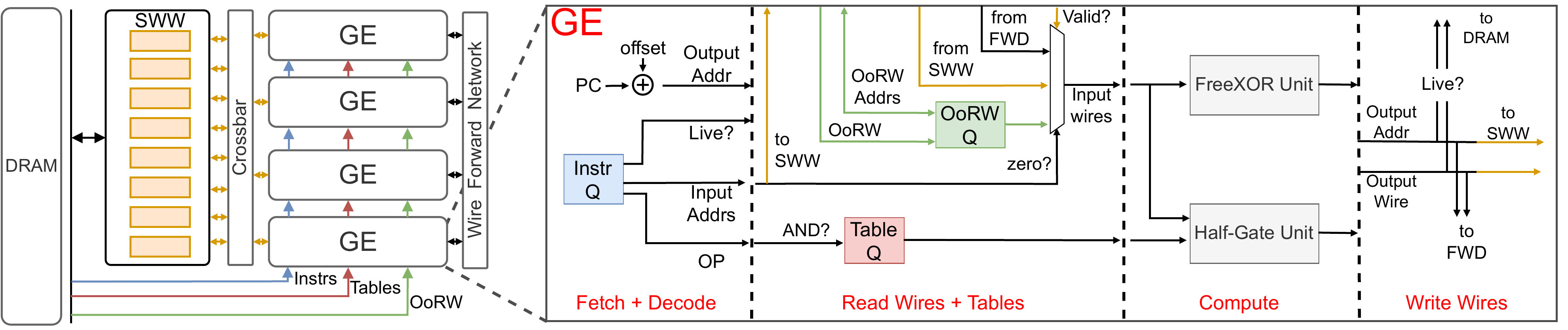}}
\vspace{-.3em}
\caption{HAAC accelerator (left) block diagram assuming four GEs and eight SWW banks.
A GE is called out (right) to show the computational pipeline.
}
\label{fig:high_arch}
\vspace{-1.em}
\end{figure*}

\textbf{Secret-sharing} enables secure computation by splitting data
into shares and is information theoretically (I.T.) secure.
Each party, of which there can be two or more, computes a function on their share 
and the results can be combined to reveal the output
\cite{secret_sharing_n_parties}.
Since both parties must work together to perform the computation, 
SS provides control over data use and confidentiality. 
A benefit of SS is that most of the costly operations can be moved offline, 
and online overheads are relatively low.
Recent research has shown these protocols work well for private neural inference, e.g., DELPHI~\cite{244032Delphi}.
The drawback is it typically relies on other PPCs in the offline phase (not Alone), e.g., using HE to compute secrets~\cite{minionn, 244032Delphi}, increasing scheme complexity and overhead.
SS, like HE, is limited in that it supports addition and multiplication, typically relying on other PPCs for non-linear functions.
Binary constructions exist~\cite{binarySS} but require many rounds of communication per function evaluation~\cite{inproceedingsABY} and have received less attention.

\textbf{Garbled circuits} are presented in detail above, here we focus on their strengths and weaknesses.
The strengths of GCs are that they support confidential computing,
control over how data is used, and processing arbitrary functions.
Others have demonstrated GCs usefulness in finance~\cite{GC_Financial}, auctions~\cite{GC_Auction}, and randomized controlled trials~\cite{GC_RandomTrial}.
Companies including Meta~\cite{fbpcf2022}, Galois~\cite{carmer2019swanky}, HP~\cite{malkhi2004fairplay}, CSIRO~\cite{MP-SPDZ} and Visa~\cite{SecureML} have used or worked on GCs.
GCs security is not based on noise, which simplifies programming 
as the user does not have to manage noise growth, as in HE.
The drawbacks include high computational overheads and large data footprints (storage and communication) due to the wires and tables.
GCs also involve both parties to take part in the computation, 
unlike in HE where only the cloud evaluates functions.


\section{HAAC Hardware}
\label{sec:hardware}

Speeding up GCs requires high-performance hardware tailored to the workload's needs.
HAAC attempts to maximize performance while minimizing hardware complexity and area.
To achieve this, HAAC takes a VLIW-inspired approach by pushing instruction scheduling, data layout, and off-chip movement management to the compiler.
In this section we present the hardware design.

Figure~\ref{fig:high_arch} (left) shows an overview of the proposed hardware.
Gate engines (GEs) are novel computational pipelines used to execute HAAC instructions (gates).
The number of GEs is configurable to the desired computational throughput.
Each GE contains a local instruction memory (streaming), table memory (streaming), 
and out-of-range wire memory (streaming).
A wire forwarding network handles intra- and inter-GE data hazards for fast resolution.
GEs share a global on-chip wire memory, the sliding wire window, which each can read and write.

The memory structures and GEs are controlled independently to
overlap all off-chip data movement with execution in a decoupled fashion.
From the GEs' perspective everything needed is on-chip,
it has no knowledge of off-chip events.
The four memory structures are controlled separately from the GEs using simple controllers configured by the compiler.
The details of how software manages control and creates streams are covered in Section~\ref{sec:compiler}.
This section details the memory subsystem and compute logic that realizes this design philosophy.

\subsection{Memory Subsystem}
HAAC's on-chip memory subsystem allocates unique structures to each GCs data type.
Distinct memories provide two benefits; 
increased parallelism: they can be accessed simultaneously, and 
improved efficiency: some structures are streaming while others require random access.

\subsubsection{Sliding Wire Window (SWW) Memory}
Wires are the inputs and outputs of gates (labels are the values of the wires).
HAAC stores wires on-chip using a scratchpad memory.
A scratchpad provides the random access support needed for input operands 
and enables HAAC to capture wire reuse across gates within a finite, contiguous wire address range.
An alternative is to stream all wires to GEs.
Streaming saves chip area by eliminating the wire SRAM and crossbar
but misses significant wire reuse opportunities.
We observe most generated wires are used by instructions that closely follow.

The wire memory is named the sliding wire window (SWW) to reflect how it is managed.
To provide random access support without address tagging,
the SWW always holds a contiguous region of wire addresses.
Assuming the wire memory can hold $n$ wires, the initial range of addresses is [0, $n-$1].
(To keep management simple, the SWW is logically partitioned in half.)
Part of the HAAC co-design is to generate output wires in a sequential address order, 
see \textit{renaming} in Section~\ref{sec:compiler}.
As the frontier of computed output wires advances past the limit of the address range
the range the SWW holds increases.
When an output wire exceeding the SWW range is generated ($n$ exceeding $n-$1), 
the SWW address range is assumed to cover a new range by remapping the first half of the space
to the next set of contiguous addresses 
(e.g., the entire SWW moves from [0, $n-1$] to [0.5$n$,1.5$n-1$])
so that the new range can capture upcoming wires.
In this way, the SWW \emph{slides} over the entire wire address space tracking output wire addresses.
When an input wire is read within the held SWW range,
it is accessed from the SWW, saving off-chip bandwidth.
Our renaming pass ensures that all wire addresses 
are properly mapped to the physically addressable 0 to $n-$1 range of the SWW.

\subsubsection{Table Memory}
Each AND gate (instruction) is associated with a unique table (constant).
The Garbler generates tables and the Evaluator consumes them, without reuse.
Each instruction is executed once, and we know each GE's instruction order at compile time.
Therefore, to optimize tables we stream them from/to each GE.
For each AND instruction, the (Evaluator) GE 
pops a table off the table queue and uses it.
The strict, known ordering of AND gates further simplifies instruction encoding, 
as table accesses are implicit and do not require addressing.

\subsubsection{Instruction Memory}
Instructions are streamed to each GE.
Queues work well as there is no control flow in instructions.
Note that GCs support conditional statements, 
and they are reflected in the circuit itself. 
Thus, there is no control flow in a HAAC program.
Therefore, random access is wasteful as instructions are sequential. 

\textit{Instruction Encoding:} 
Each HAAC instruction specifies the gate's operation (2b), 
two input wire addresses (17b each for 2 MB SWW), 
and if the output wire is \emph{live} (1b) after its SWW window,
needing to spill to DRAM.
Wire output addresses are not specified as they are generated in-order,
see renaming in Section~\ref{sec:compiler}.
Computing the addresses using the instruction's program position saves encoding space.

\subsubsection{Out-of-Range Wires}
The SWW filters most wire accesses. 
However, GCs support arbitrary logic and inevitably wire accesses will
exceed the range currently held on-chip.
HAAC leverages two properties to avoid the drawbacks of
a standard cache or pull-based design.
It is known when and which wire accesses will be out-of-range (OoR),
eliminating the need to check the SWW (it is not there) and
the need to rely on a pull-based access event, 
which would introduce costly stalls into HAAC's in-order pipeline.
To optimize for this HAAC implements a third, GE-local queue named the out-of-range wire (OoRW) queue.
The head of this queue contains the wire needed by the next instruction incurring an OoR access, which the compiler can determine.
The zero wire address 
in an instruction
is reserved to indicate OoR and that the wire should be read from the queue, not the SWW.
If both operands are OoR, the first operand is handled first.

OoR wires are addressed using 32 bits.
These addresses are streamed on-chip (in-order) and used to fetch OoR wires.
To guarantee correctness, HAAC also uses a valid bit for wires.
When an OoR wire is read from DRAM, 
its
valid bit is checked and wires are not pushed to the OoRW queue until a valid read is performed.
Reading invalid OoR wires from DRAM is inefficient, but it rarely impacts performance.
When a live wire is to be written back to DRAM
it also remains valid on-chip 
(in the SWW) 
for at least the time it takes to process 
instructions proportional to half of the SWW size.
For example, the SWW wire range [0.5$n$, $n-1$] is valid for all instructions associated with SWW range [0.5$n$, 1.5$n-1$] and only becomes OoR when moving to process the wire range [$n$, 2$n-1$].
Therefore, there is ample time between live wires being written to DRAM (which initiates as soon as they are computed) and requested by GE (through the OoRW queue) much later.

By preemptively pushing all OoR accesses to the OoRW queue,
HAAC is able to eliminate all pull-induced long-latency access events,
enabling the conversion of all off-chip HAAC data movement to streaming.
Pushing OoR wire reads to the queue is the key to enabling complete decoupling between data movement and execution in HAAC hardware;
it is enabled by HAAC's co-design approach.

\subsection{GE Pipeline}

The GE pipeline, Figure~\ref{fig:high_arch} (right), constitutes a simple frontend for fetching and decoding instructions, stages to read wires and tables, execution units for Half-Gate (AND) and FreeXOR (XOR), write-back, and forwarding logic.
GEs are deeply pipelined to run at high frequency and overlap instructions, leveraging workload ILP.

\begin{figure*}[t]
\centerline{\includegraphics[width=.93\textwidth]{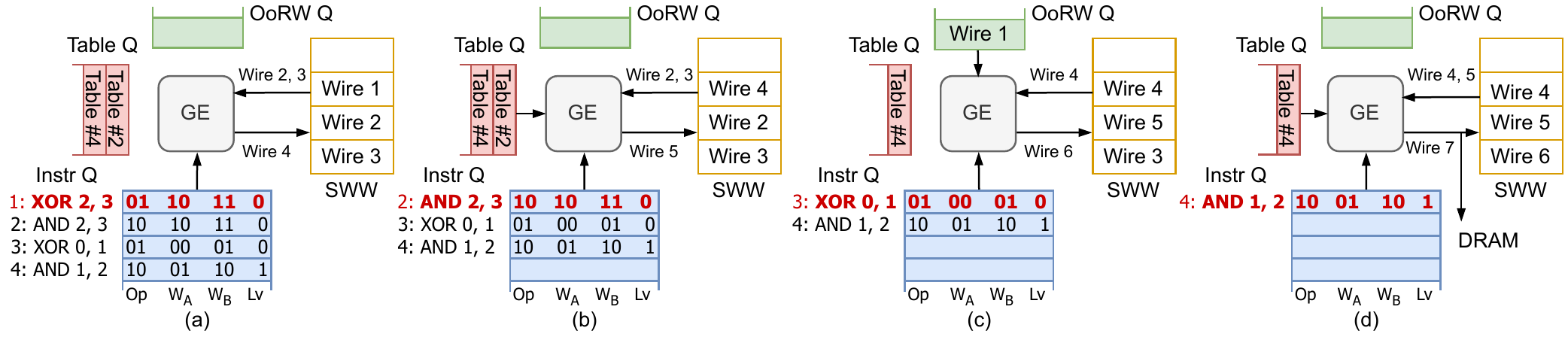}}
\vspace{-.5em}
\caption{
An example of HAAC (Evaluator) processing instructions.
(a) shows a XOR gate using wires from the SWW;
(b) illustrates the Table queue by processing an AND gate;
(c) depicts the reading of an OoRW through the OoRW queue, using 0 as the wire address;
(d) shows a live wire being written back to DRAM.
}
\label{fig:running_example}
\vspace{-.5em}
\end{figure*}

\begin{figure*}[t]
\centerline{\includegraphics[width=.9\textwidth]{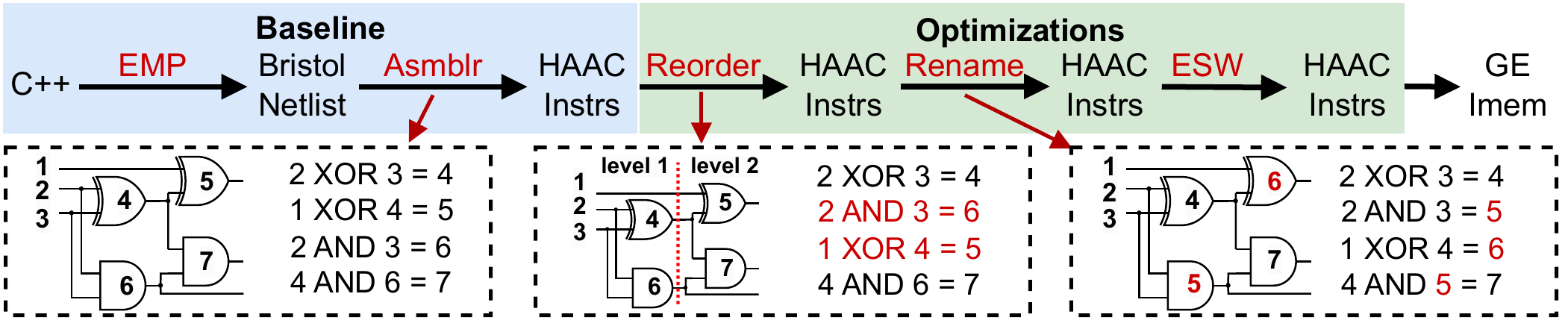}}
 \vspace{-.5em}
\caption{
An overview of the HAAC compiler flow and illustration of optimizations.
Reordering optimizes instruction schedules to improve parallel execution and
renaming linearizes output wire addresses to match program order, which increases the effectiveness of the SWW.
}
\label{fig:compiler}
\vspace{-1em}
\end{figure*}

\textbf{Frontend:}
Fetch and decode logic are simple as no control flow nor memory instructions are needed, as mentioned above.
The fetch and decode stages fetch the next instruction off the instruction queue, determine which of the three instruction types (AND, XOR, nop) it is, compute the output wire address wire address and forward addresses to proceeding stages.

\textbf{Read Wires and Table:} Wire addresses are used to index the SWW.
Reads are split across three stages, and a crossbar interfaces SWW banks with GEs.
It takes one cycle to get the address to the bank, one to read a bank, and one to get data from the bank to the GE.
Each stored wire includes a valid bit to indicate the value has been computed.
If false, the wire label is being computed in a GE and it will be retrieved via the forwarding network.
When a zero wire address is decoded the wire is accessed from the OoR wire queue.
For AND instructions, a table is retrieved from the head of the table queue in parallel with the input wire accesses.

\textbf{Compute Units:} The heart of the GE is the execution pipeline.
To maximize performance, we developed custom logic units
for processing Half-Gate and FreeXOR computations.
In GCs, a party can be either a Garbler or Evaluator, needing support for only one.
HAAC includes distinct compute units for the Garbler and Evaluator while the rest of the pipeline is shared.
Each unit was designed using High-Level Synthesis (HLS) with the EMP~\cite{emp-toolkit} GCs framework as a reference and to validate correctness.

\textbf{Half-Gate Unit:}
Native EMP code is not amenable to high performance HLS programming.
The main issues stem from key expansion and AES modules. 
Both modules execute the same function many times, sequentially.
Many arrays in the design (e.g., AES keys) were initially implemented as SRAMs, 
which are more area efficient than flip-flops but restrict data access per cycle.
Further, allocated hardware was re-used across different modules,
e.g., the S-BOX lookup table is implemented as a single ROM. 
With one S-BOX instance, only a single round of key expansion \emph{or} AES could occur at a time.
(These SRAMs are embedded in the pipeline and distinct from wire/table SRAMs, 
accesses are implicit.)
The reuse of these structures limits parallelism and the ability to pipeline the long-latency computation,
resulting in low throughput and clock frequency.

To improve performance, we replicated reused round structures 
(e.g., S-BOX, MixColumns, and XOR arrays) 
with the inline HLS pragma to alleviate resource contention.
All SRAM instances within the units were flattened to registers, 
enabling direct access to different array elements at the same time,
providing multiple data accesses per cycle.
Manual code rewriting was done to
resolve false dependencies,
unroll loops, and explicitly instantiate parallel logic
HLS could not find.
The remaining loops were unrolled with pragmas.
Our optimized design is fully pipelined, capable of accepting a new input
every cycle.
The Garbler and Evaluator GE have 21 and 18 stage pipelines
for Half-Gate,
respectively.
No changes impact correctness, which we verified against EMP Toolkit.

\textbf{FreeXOR Unit:}
The FreeXOR unit is much simpler than the Half-Gate.
It is implemented using an array of XORs and takes a single cycle.
XOR hardware exists in the AES logic used in the Half-Gate unit but
is not shared to allow FreeXORs to run in parallel.
The benefit is that XORs can complete in one cycle and immediately resolve dependencies rather than incurring the full Half-Gate pipeline latency.

\textbf{Write and Forward Wires:}
As computations finish, the output wire label is written back to SWW (two cycles)
and, if the live bit is set, to DRAM.
GEs also support forwarding (FWD).
When there is a wire address match between a completing instruction and one in the frontend,
the writeback stage forwards the wire's values.
When multiple GEs are used the forwarding logic extends across GEs
and the overall area is a function of the number of GEs used.
In practice, we find the logic is not expensive, taking only 0.002mm$^2$ in a 16 GEs design.
We could encode the data dependence information in the ISA
(e.g., like some CGRA or Dataflow machines)
but this complicates the compiler, frontend, and increases instruction size.
As the forwarding logic is simple and inexpensive, we use it in HAAC.

\subsection{Example}
An example of HAAC (Evaluator) processing instructions with a very small SWW is shown in Figure~\ref{fig:running_example}.
Note the instructions are the same as the compiler example in Figure~\ref{fig:compiler} for continuity.
When starting, input wires (1, 2, 3) are loaded to on-chip memory (SWW).
(a) shows an XOR gate where both wires (2, 3) are in the SWW, 
the output wire 4 is 
written to SWW entry 1, overwriting wire 1.
(b) depicts an AND gate.
A table is read from the head of the Table queue,
both input wires (2, 3) are in the SWW, and
the output wire 5 is stored to SWW entry 2.
Writing wire 5 to SWW entry 2 overwrites wire 2 and advances the address range of the SWW to [3-5].
Overwriting is safe as any wires still needed spill to DRAM as indicated by the live bit.
In (c), wire 1 of the XOR instruction is not in the SWW and is marked as 
0
by the compiler. 
The GE knows wire 1 is in the head of the OoRW queue and reads it.
Finally, in (d) an AND executes where all operands are in the SWW and a table is retrieved from Table queue. 
The output wire 7 is marked as live (Lv=1) and is saved to DRAM as it will be needed later by the program. 
Note that SWW entry 0 is not used as the address is reserved for the OoRW queue.
In practice, SWWs have tens of thousands of entries, and one slot is negligible.


\section{HAAC Compiler}
\label{sec:compiler}


This section presents the HAAC compiler.
The compiler has two jobs: optimize programs for high-performance and generate queue streams.
We begin by presenting the overall compiler flow and then detail three optimizations for performance.

\subsection{Overview}
The software workflow is shown in the top of Figure~\ref{fig:compiler} and proceeds as follows.
(We use the same instructions as Figure~\ref{fig:running_example} for a complete example).
First, a program, written in C++, is input to the EMP Toolkit~\cite{emp-toolkit}.
The GCs framework is widely used and provides high-level programming support.
EMP analyzes programs and outputs a netlist in Bristol format~\cite{Bristol}.
These netlists are input to our HAAC assembler, which outputs HAAC instructions.
The output from the assembler is a baseline HAAC program to which optimizations are compared.
Optimization details are provided below and are applied in the order they are introduced.
Reordering (RO) optimizes instruction schedules, renaming (RN) linearizes gate output wire addresses, and eliminating spent wires (ESW) elides redundant writes to off-chip memory.

The final step of the compiler is to generate queue streams.
All queues are GE-local, and the first step is to determine which instructions are processed in each GE.
This is done by mapping instructions from the program to non-stalled GEs each cycle in our simulator,
saving the order, and replaying it in hardware.
This mitigates load imbalance and eliminates the need for expensive super-scalar-like hardware.
Next, knowing the order of instructions enables the compiler to determine the table order by inspecting each GE instruction stream.
Lastly, the order of OoR wire accesses must be determined.
This is done by comparing all instruction input wires against the range of wires currently held in the SWW.
Each time an OoR wire is encountered, its address is appended to the queue and used to fill the OoR wire queue.
During this pass, the input wire operands with OoR wires are replaced with zero.

\subsection{Optimizations}

Baseline HAAC programs are built directly from EMP by converting a list of gates into HAAC instructions.
However, these tend to perform poorly as the programs do not consider HAAC's hardware, leaving potential performance unrealized.
In this section we present three optimizations for better instruction scheduling (reordering), improved on-chip reuse (renaming), and fewer writes to off-chip memory (eliminating spent wires),
as shown in Figure~\ref{fig:compiler}.

\subsubsection{Exploiting ILP With Reordering}
GCs programs generally have high ILP, but parallelism is lost in the baseline program as instructions are scheduled
following a depth-first circuit traversal, i.e., 
in tight producer-consumer relationships minimizing the distance between dependent gates.
A depth-first traversal can save off-chip traffic, as wires are reused across neighboring instructions, however it generally restricts parallelism.
We empirically find this results in a significant number of stalls in GEs as they are in-order and must wait for dependence to resolve.
Performance can be improved by scheduling HAAC instructions with more distance between dependencies.
We propose two scheduling schemes.

\textbf{Full Reorder:} To maximize instruction parallelism,
we first order HAAC programs according to their ILP level order (breadth-first).
To do this, we build a leveled dependence graph of the entire HAAC program, exposing all available ILP.
Next, we iterate through each level one node (instruction) at a time,
appending traversed nodes to a new instruction list.
This approach maximizes program parallelism
since all instructions within a level are independent.
A full reordering works well for resolving data hazards but can increase DRAM traffic due to less wire reuse in the SWW.
E.g., if a level contains too many gates and their output wires exceed the capacity of the SWW, they spill to DRAM.
Strictly prioritizing instruction parallelism can result in missed opportunities for input wire reuse while benefiting compute parallelism.

\textbf{Segment Reorder:} 
To better balance wire reuse and improve instruction parallelism we developed segmented reordering.
Here, rather than computing the ILP graph for the entire program
we partition the baseline program along the depth direction into contiguous parts (segments) and reorder instructions within each segment.
We set the segment size to half the size of the SWW,
as this is how the SWW is logically partitioned.
Doing so keeps segments large (e.g., 65,536 instructions per segment for a 2 MB SWW)
and the ILP within segments is usually sufficient to avoid stalls.
By restricting segments to half the SWW range,
we preserve wire locality from the baseline
and can capture it in the SWW.
Segment reorder provides more instruction parallelism than baseline programs 
and generally captures more wire reuse than full reordering.

\subsubsection{Linearizing Output Wires With Renaming}
The SWW provides on-chip wire accesses for a contiguous portion of the wire address space.
Without structure, especially after reordering, there is no meaningful correlation between the program order and wire addresses used in each instruction.
To effectively utilize the SWW we rename the \emph{output} wires of each (post-reorder)
instruction to follow the program order, and then propagate address mapping changes to the input wires as in Figure~\ref{fig:compiler}.
Renaming has two advantages.
First, it concentrates wire address accesses to the range currently supported by the SWW.
By linearizing output wire addresses to match the instruction order we optimize for wires being generated, stored on-chip, and reused.
Second, linearizing outputs saves instruction encoding space as output wire addresses are incremental.

\begin{table}[t!]
\centering
\caption{
Key characteristics of the benchmarks used.
Levels indicate circuit depth and Spent wires assume a 2MB SWW with full reordering.
}
\label{tab:benchmarks}
\resizebox{\columnwidth}{!}
{
\setlength{\tabcolsep}{1mm}{
\begin{tabular}{ccccccc}
\hline
\textbf{Benchmarks} & \textbf{\# Levels} & \textbf{\# Wires (k)} & \textbf{\# Gates (k)} & \textbf{AND \%} & \textbf{ILP} & \textbf{Spent Wire \%} \\ \hline
BubbSt              & 75636              & 12542                 & 12534                 & 33.33           & 166          & 99.87                \\
DotProd             & 277                & 389                   & 381                   & 34.39           & 1376         & 86.43                \\
Merse               & 1764               & 1444                  & 1444                  & 27.15           & 818          & 98.49             \\
Triangle            & 1403               & 6984                  & 6979                  & 34.02           & 4974         & 56.76                \\
Hamm                & 76                 & 410                   & 328                   & 25.00           & 4311         & 99.93                \\
MatMult             & 157                & 1519                  & 1515                  & 34.48           & 9649         & 82.16                \\
ReLU                & 2                  & 133                   & 68                    & 96.97           & 33792        & 49.23               \\
GradDesc            & 106314             & 6344                  & 6343               & 42.91           & 60           & 99.70                \\ \hline
\end{tabular}
}
}
\vspace{-1.5em}
\end{table}

\subsubsection{Saving Write Bandwidth With Eliminating Spent Wires (ESW)}
Not all computed wires need to be written back to DRAM.
Each generated wire is used a finite number of times,
and in many cases, wires are only ever reused through the SWW.
The HAAC compiler flags wires that need to be written 
to DRAM by setting the live bit in the instruction;
\emph{spent} wires are those not needed by future instructions beyond their available SWW wire range.
This is implemented in the compiler by
checking whether an output wire is ever used past its current SWW boundary.
Live wires stored to DRAM are brought back as OoR wires through the OoRW queue as needed.
ESW is highly effective;
using a 2MB SWW, on average only 16\% of wires are live and written off-chip, see Table~\ref{tab:benchmarks}.

\section{Methodology}
\label{sec:methodology}
In this section we detail our experimental setup.
CPU performance is measured on an Intel Core i7-10700K~\cite{corei7} running at 3.80GHz.
CPU power is collected using a commercial tool~\cite{hwinfo_2021}.
We use the EMP Toolkit~\cite{emp-toolkit, mpc-sok} as our software framework.
EMP leverages AES-NI~\cite{AESNI} for high performance and a competitive baseline.

\textit{Simulator:}
We developed a cycle-accurate simulator to evaluate HAAC and explore design tradeoffs.
The simulator is built as two parts: GEs and memory. 
We model GEs using our hardware implementation of Garbler/Evaluator logic described in Section~\ref{sec:hardware}.
HAAC uses multiple clock domains, 1 GHz and 2 GHz for the GEs and SWW, respectively.
The SWW is implemented as a collection of single-port SRAM banks.
Each SRAM word stores a wire label and valid bit to indicate whether data is ready to use.
We empirically evaluate how SWW banks and GEs interact and find that 4 banks per GE works well to minimize banking (area overhead from partitioning) while avoiding contention; we use this ratio in our evaluation.
We evaluate HAAC using two types of DRAM:
a 
DDR4-4400 at 35.2 GB/s~\cite{DDR4, crucial_dram_speed} and a HBM2 PHY at 512 GB/s bandwidth, 
as reported in~\cite{feldmann2021f1, intel_fpga_hbm, a100hbm2, RambusHBM2E}.
The simulator is verified to be functionally correct and handles stalls by 
precisely tracking all data movement through the machine.

\textit{Benchmarks:}
We evaluate HAAC using 8 benchmarks from VIP-Bench~\cite{vipbench}, 
see Table~\ref{tab:benchmarks}.
Benchmarks were selected to be representative as characterized in the VIP-Bench paper:
shallow, deep, complex, and simple.
To evaluate the performance of HAAC and consider relevant problems,
we either use the original data sizes or scale up input sizes to better stress the hardware and off-chip behavior as some are too small.
We increase the size of Dot Product to two 128 elements 32-bit integer vectors,
Matrix Multiplication to 8$\times$8 integer matrices,
Hamming Distance to 40960-bit length,
and ReLU to be evaluated 2048 times.
Linear Regression uses 20 rounds of Gradient Descent and is implemented with true floating point arithmetic.
Related work does not use VIP-Bench 
which was released after the papers were published.
Workloads reported in prior work are \emph{much} smaller than VIP-Bench, 
and we only report their performance for comparison.
For example, the 8-bit Millionaire-Problem benchmark used in FASE~\cite{8735500} has only \emph{33 gates}, 
the smallest VIP-Bench workload has 68 thousand.

\textit{CAD Tools and Technology:}
We used Vivado HLS 2018.3~\cite{xilinx2020vivado} to implement the Half-Gate unit.
Hardware was synthesized without FPGA IP as Verilog.
The forwarding network, crossbar, and pipeline stages were hand implemented in Verilog using Vivado Design Suite 2018.3. 
We used TSMC 28HPC as our technology node~\cite{TSMC28HPC}.
Verilog logic was synthesized using Design Compiler (Version T-2022.03)~\cite{SynopsysDC}.
SRAM power and area numbers are from the TSMC N28HPC+ Memory Compiler~\cite{TSMCSRAMcompiler}.
The synthesized netlist was placed and routed using Cadence Innovus 18.1~\cite{Innovus}.
The layout was designed to have a utilization of 70\% before place-and-route;
power and area numbers were extracted after timing was met.
To match prior work on high-performance PPC hardware we scale our memory and standard cell based logic structures from 28nm to 16nm using foundry provided scaling factors~\cite{28nmto16nm, 20nm}.
These reduce 28nm power by 60\% and area by 1.9$\times$.

\begin{figure}[t!]
\centerline{\includegraphics[width=.95\columnwidth]{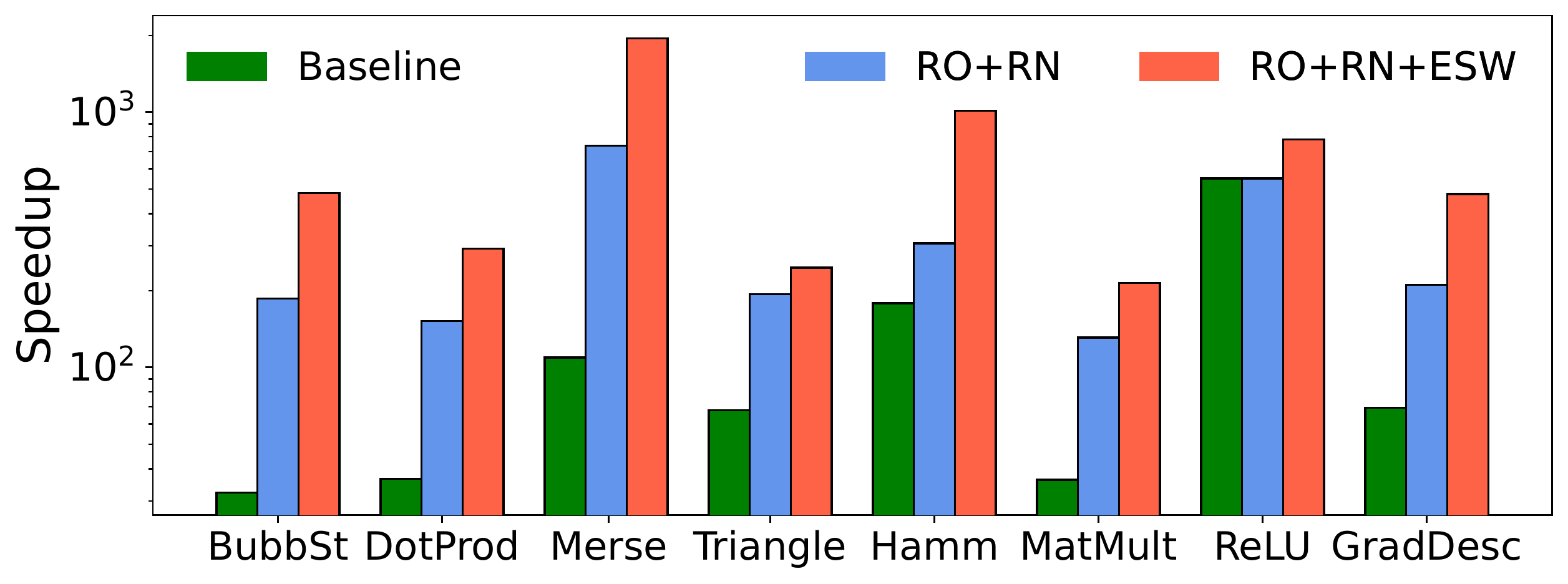}}
\vspace{-.5em}
\caption{HAAC speedup over a CPU.
Results are shown for an original schedule (Baseline), 
full reordered and renamed (RO+RN), 
full reordered and renamed with eliminating spent wires (RO+RN+ESW).
}
\vspace{-1em}
\label{fig:compiler_optimizations}
\end{figure}

\textit{Correctness:} 
Correctness is vital; we validate at each step and 
all designs were functionally verified against EMP.
Ported VIP-Bench workloads are tested using multiple inputs. 
We instrumented EMP to dump input-output pairs of components (e.g., the Half-Gate and FreeXOR) to test C++ HLS code. 
We then tested EMP and C++ HLS code with hundreds of I/O pairs; all matched perfectly. 
RTL was tested by simulating each block using the same test vectors.

\section{Evaluation}
\label{sec:evaluation}

In this section we evaluate HAAC using the simulator and benchmarks from Section~\ref{sec:methodology}.
We demonstrate HAAC's performance,
the strength of the compiler optimizations,
characterize area and energy,
and compare against prior work.

\subsection{Compiler Optimizations}

We begin our evaluation by showing the efficacy of the HAAC compiler.
This experiment assumes 
an Evaluator HAAC
with 16 GEs, a 2MB SWW, and DDR4.
Figure~\ref{fig:compiler_optimizations} shows speedup results relative to the original EMP program (baseline) running on the CPU.
Green bars indicate HAAC's speedup using baseline instruction schedules.
Blue bars show the performance of executing the same program after running full reordering and renaming optimizations. 
Red bars include eliminating spent wires (ESW) and are discussed next.

Starting with the baseline bars (green), we make two observations.
First, the results show that even with a baseline program HAAC obtains good results,
with an average speedup of 82.6$\times$.
This demonstrates the effectiveness of the GE design over the CPU.
Next, after the compiler fully reorders and renames
the baseline program (blue bars)
we obtain an additional average speedup of 3.1$\times$ over the baseline,
showing the effectiveness of the technique to increase parallelism and better utilize GEs.
We group reordering and renaming as without renaming the SWW is ineffectual.
Reordering alone causes a program's wire accesses to span the wire address space and increases off-chip wire traffic.
Therefore, renaming is run by default after any reordering in the remainder of the text.
Across all benchmarks we find that the maximum reordering and renaming speedup is 6.8$\times$ in the Mersenne-Twister benchmark.
Full reordering and renaming does not speed up ReLU.
Table~\ref{tab:benchmarks} shows ReLU circuits have only two dependence levels, 
and the baseline program already contains significant parallelism.
Reordered programs are fast enough to saturate DDR4 bandwidth,
our next optimization shows how we reduce bandwidth pressure.

We note that on a CPU, garbling is 11.9\% slower than evaluation, but the HAAC Garbler is only 0.67\% slower than the HAAC Evaluator, averaged across all benchmarks.
Therefore, we expect a higher HAAC speedup for Garbler and conservatively report Evaluator speedup here.  


\begin{table}[t!]
\caption{
Comparison of wire traffic between segment and full reordering 
(both use ESW).
Live, OoRW, and total are shown as the total number of kilo (k) wires assuming a 2MB SWW. 
Top benchmarks favor segment and bottom full reordering.
}
\vspace{-.5em}
\label{tab:seg_benefit_table}
\resizebox{\columnwidth}{!}{
\centering
\setlength{\tabcolsep}{3mm}{
\begin{tabular}{ccccc|cc}
\hline
         & \multicolumn{2}{c}{Live Wires (k)} & \multicolumn{2}{c|}{OoRW (k)} & \multicolumn{2}{c}{Total (k)} \\
         & Seg              & Full            & Seg           & Full          & Seg           & Full          \\ \hline
MatMult  & 6.01             & 271             & 495           & 582           & \textbf{501}  & 853           \\
DotProd  & 5.59             & 52.8            & 91.5          & 56.8          & \textbf{97.1} & 110           \\
Merse    & 0.06             & 21.8            & 0.05          & 29.4          & \textbf{0.11} & 51.2          \\
Triangle & 52.4             & 3020            & 2411          & 5934          & \textbf{2463} & 8954          \\
ReLU     & 67.5             & 67.6            & 2.11          & 2.05          & 69.6          & 69.7          \\ \hline
BubbSt   & 161              & 16.6            & 750           & 37.2          & 911           & \textbf{53.8} \\
GradDesc & 17.3             & 19.2            & 392           & 344           & 409           & \textbf{363}  \\
Hamm     & 0.75             & 0.27            & 1.22          & 0.26          & 1.97          & \textbf{0.53} \\ \hline
\end{tabular}
}
}
\vspace{-1.5em}
\end{table}

\subsection{Optimizing Off-Chip Memory Traffic}

\textit{Eliminating Spent Wires (ESW):}
We now analyze the performance benefits of ESW.
The red bar in Figure~\ref{fig:compiler_optimizations} indicates the performance of full reordering with ESW.
Table~\ref{tab:benchmarks} shows by adding the live bit to instructions
we can save an average of 84\% of wires from being stored back to off-chip memory, freeing up significant bandwidth.
By reducing the wires that need to be stored off-chip we achieve an average of 2.1$\times$ more speedup than full reordering and renaming alone.
Hamming-Distance shows the most speedup, 3.3$\times$,
which matches expectations from Table~\ref{tab:benchmarks}.
The performance gain from ESW also indicates that the reordered programs are memory bound, and thus ESW provides speedup over full reordering.

\textit{Segment Reorder:}
Full reordering improves parallelism but can spread wire accesses over too wide a range within a limited instruction window, this can prevent the SWW from capturing wire reuse.
Segment reordering is a compromise between the baseline and full reordering.
Table~\ref{tab:seg_benefit_table} compares the wire traffic of the two approaches.
We note that baseline and segment 
reordering have almost
the same traffic and omit baseline for space.
Columns delineate 
storing live and reading OoRW traffic with the total on the right.
Different reordering schemes do not impact ReLU's wire traffic. 
It computes 2048 independent ReLUs, reflective of real-world workloads.
Because each ReLU is not reused and results are stored off-chip, wire traffic does not change much.
Benchmarks on the top of the table (MatMult to ReLU) are receptive to segment reordering,
reducing wire traffic by 5.0$\times$ (geomean) compared to full reorder.
Those in the bottom (BubbSt, GradDesc, Hamm) favor full reordering.
To understand why we detail the behaviors of two representative benchmarks from each group.


\begin{figure}[t!]
\centerline{\includegraphics[width=.95\columnwidth]{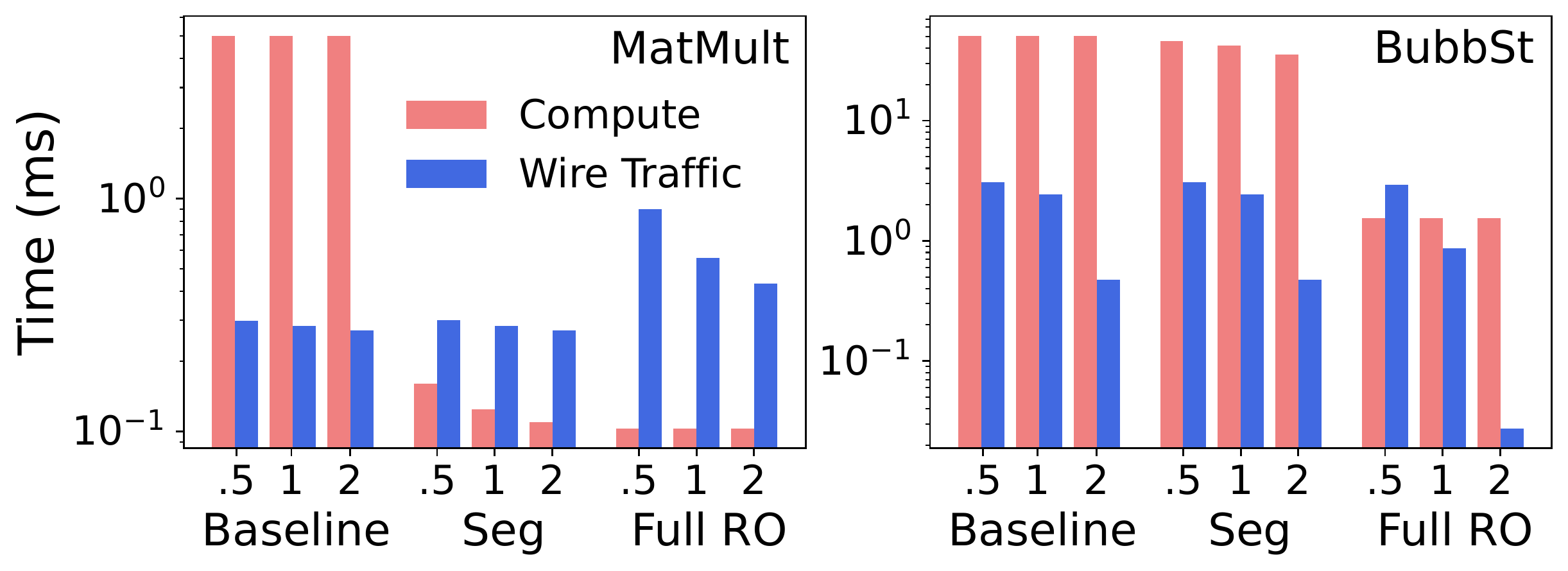}}
\vspace{-1em}
\caption{Performance of three instruction orders (Baseline, Segment, Full Reordering) for MatMult and BubbSt. 
Compute isolates GE execution time whereas wire traffic shows only off-chip wire data movement time.
X-ticks (.5, 1, 2) indicate SWW size (MB).
}
\vspace{-1.6em}
\label{fig:segment}
\end{figure}

Figure~\ref{fig:segment} analyzes the performance of different orderings using two representative benchmarks: Matrix Multiplication (MatMult) and Bubble Sort (BubbSt).
The two bars show the compute (no off-chip latency, red) and off-chip wire traffic (no compute latency, blue) times for the baseline, segment, and fully reordered program.
Overall performance is constrained by the higher bar.
Each ordering is evaluated using three SWW sizes (0.5, 1, and 2 MB). 
We set the segment size to half the SWW size, which we find performs best,
and again assume an accelerator with 16 GEs and DDR4.

Increasing the SWW size reduces wire traffic time (blue), as expected.
A larger SWW covers a wider wire address range and increases on-chip wire reuse,
reducing wires read as OoR and saved as live.
Increasing SWW size does not change the compute time for the baseline or fully reordered programs.
Since the segment size is half of the SWW size, 
segment compute time can improve as more parallelism is found in a larger window.

Segment reordering is highly effective for MatMult. 
As Figure~\ref{fig:segment} shows, MatMul is compute bound in the baseline.
Full reordering improves compute performance by 48.8$\times$,
unlocking the parallelism in the program.
However, it does so at the expense of wire memory traffic, 
reducing on-chip reuse and increasing traffic time by 2.0$\times$ for a 1MB SWW.
This is because MatMult has high ILP, and a strictly breadth-first traversal of the graph overwhelms the limited capacity of the SWW, increasing wire traffic.
Segment reordering balances compute parallelism and memory bandwidth.
By restricting reordering to a program segment, it shows similar wire traffic as the baseline
while still improving compute time through increased parallelism.
The wire traffic of baseline is not always optimal.
One case is BubbSt, see Figure~\ref{fig:segment}.
BubbSt has long chains of dependent instructions, large dependence fan-out (i.e., instructions from one ILP level connect to many in the next), and relatively low ILP.
The baseline program processes instructions following the order of the long dependence chains.
This limits parallelism and does make use of the SWW while also resulting in many live wires.
Wires are reused through the SWW, but because of the large fan-out, must be stored to DRAM for use again later in the program.
With segment reordering, these long chains of dependent instructions consume much of the segment window.
Compute performance is improved by reordering across a few long, independent chains, but not much.
Similarly, most wires are live and must spill to DRAM.
With full reorder, however, both compute time and wire traffic improve.
Because levels are shallow, a SWW of 2MB can fit multiple entire levels of wires.
This allows HAAC to exhaust wire reuse on-chip through the SWW and reduce live wire writebacks while maximizing compute parallelism.
In practice, we can run both and deploy the best performing optimization, as performance is deterministic.

\begin{figure}[t!]
\centerline{\includegraphics[width=.98\columnwidth]{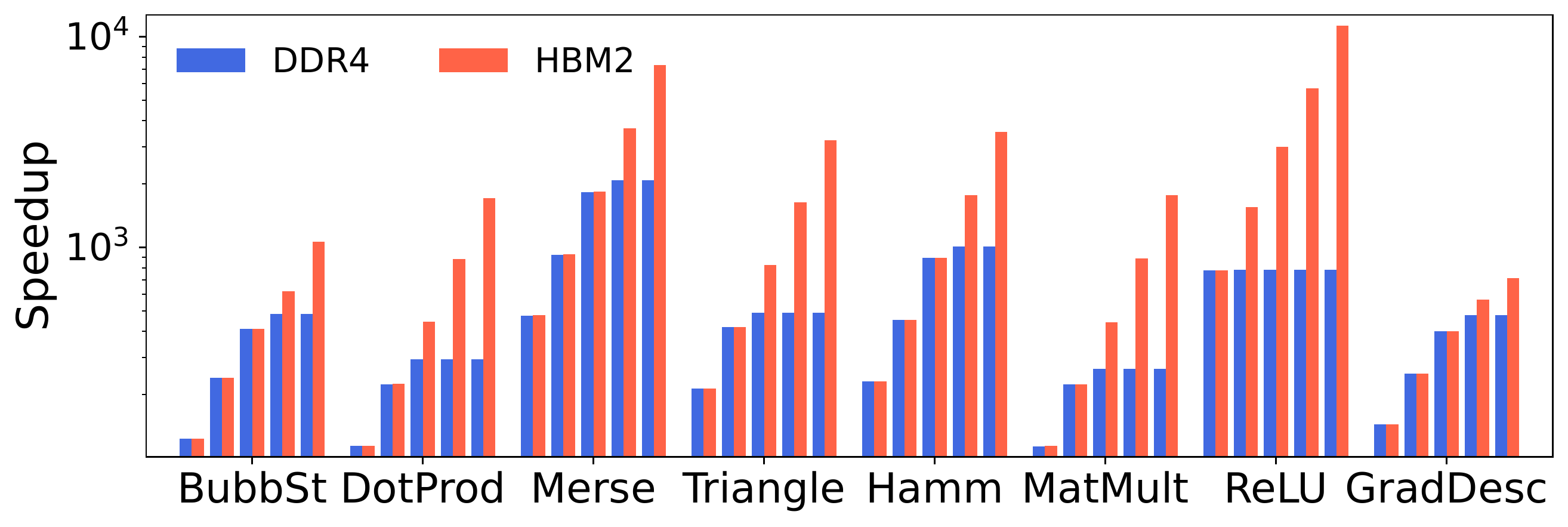}}
\vspace{-1em}
\caption{Benchmark performance scaling with respect to GE count, bar clusters show 1, 2, 4, 8, and 16 GEs and speedup is relative to the CPU.}
\vspace{-1.5em}
\label{fig:mc_speedup}
\end{figure}

\subsection{Parallel GE Execution and Speedup Over Software}

Figure~\ref{fig:mc_speedup} shows speedup relative to CPU performance.
We evaluate performance by scaling GEs from 1 to 16 and use a 2 MB SWW.
Two types of DRAM are assumed: DDR4 to fairly compare with our CPU and
HBM2 to understand how HAAC benefits from advanced memory technology.
With DDR4, benchmarks are reported using the reordering (segment or full) with better performance.
When using HBM2, all benchmarks use full reordering to maximize ILP and the utilization of available bandwidth.

We find that in most cases performance scales well when increasing the number of GEs initially, but designs can saturate DDR4 bandwidth and become memory bound.
This can be seen when blue GE speedup bars plateau.
With HBM2 the performance continues to scale across the range of GEs considered.
When a red bar is greater than its corresponding blue bar, HAAC is constrained by DDR4's memory bandwidth, and HBM2 can help continue to scale performance.
With HBM2, we find that the maximum speedup increase from one to sixteen engines is 15.5$\times$ (for MatMult) while the geomean speedup is 12.3$\times$ for the Evaluator (the Garbler is nearly identical).
In Table~\ref{tab:benchmarks}, we can see many benchmarks have substantial ILP, 
which HAAC successfully leverages to achieve near ideal speedup from 1 to 16 GEs.
Performance scaling is constrained in Gradient-Descent, which uses floating point, and Bubble Sort due to their lack of ILP.

Regardless of which design is used, we find HAAC provides substantial benefits over the CPU implementation.
Using a single GE with HBM2, a HAAC accelerator provides a maximum speedup of 779$\times$ (ReLU) and geomean of 213$\times$ across all benchmarks.
When going to the highly parallel 16 GEs design, 
we observe a geomean speedup of 2,616$\times$ and maximum speedup of 11,330$\times$ (ReLU), again assuming HBM2 and full reordering for all benchmarks.

\subsection{Area, Power, and Energy Analysis}

Table~\ref{tab:area_power_table}
shows the area and power numbers of a 16 GEs design with a 2 MB SWW and 64 SWW banks.
The design uses a 64 KB SRAM for table, instruction, and OoRW queues, 
as well as a single HBM2 PHY~\cite{feldmann2021f1, a100hbm2, RambusHBM2E}.
In HAAC, most of the chip area goes to the GE, specifically the Half-Gate and SWW.
The total area for HAAC is 4.3 mm$^2$ in 16nm.
Given the small area footprint, we assume HAAC would be used as an IP in a larger SoC.
Therefore, the PHY would be shared, and we focus on reporting HAAC IP area.
We include all numbers so users interested in standalone chips are aware of the high HBM2 PHY cost relative to HAAC's size.

\begin{table}[t!]
\centering
\caption{
A breakdown of HAAC chip area and average power.
}
\vspace{-.5em}
\label{tab:area_power_table}
\resizebox{.85\columnwidth}{!}
{
\setlength{\tabcolsep}{3mm}{
\begin{tabular}{lrr}
\hline
Component           & Area (mm$^2$) & Power (mW)    \\ \hline
Half-Gate           & 2.15          & 1253          \\
FreeXOR             & 9.51E-04      & 0.321         \\
FWD                 & 1.80E-03      & 0.255         \\
Crossbar            & 7.27E-02      & 16.6          \\
SWW (SRAM)          & 1.94          & 196           \\
Queues (SRAM)       & 0.173         & 35.5          \\
\textbf{Total HAAC} & \textbf{4.33} & \textbf{1502} \\ \hline
HBM2 PHY            & 14.9          & 225 (TDP)     \\ \hline
\end{tabular}
}
}
\vspace{-.5em}
\end{table}

\begin{figure}[t!]
\centerline{\includegraphics[width=.95\columnwidth]{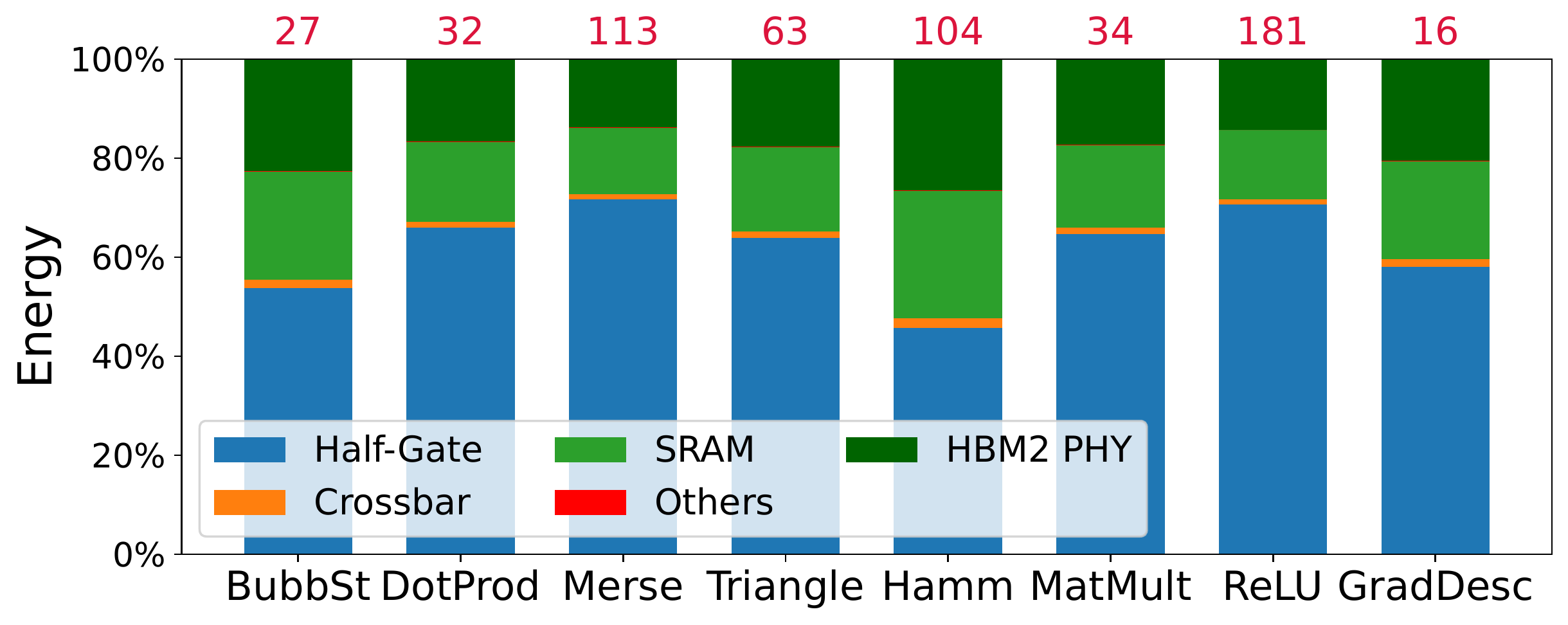}}
\vspace{-1em}
\caption{The normalized energy consumption of HAAC components.
Energy efficiency over the CPU (in K$\times$) is shown on top in red.
}
\vspace{-1em}
\label{fig:energy_bar_cpu}
\end{figure}

The average power dissipation across benchmarks is 1.50W
(1.73W including HBM2),
resulting in a power density of 0.35W/mm$^2$.
The high switching rate of the cryptographic circuits and SWW frequency contribute to the moderate power dissipation.
Figure~\ref{fig:energy_bar_cpu} shows the energy consumption breakdown for each 
full reordered
benchmark.
Note that FreeXOR
and the forwarding network are so small, they are grouped as Others.
The Half-Gate consumes most of the energy, on average 61\% among all benchmarks.
The CPU is significantly slower and higher power than HAAC,
dissipating an average of 25W across benchmarks.
The red text atop each bar indicates HAAC energy efficiency improvement over the CPU baseline.
On average, HAAC is 53,060$\times$ more energy efficient than the CPU.

\subsection{Comparing to Plaintext}
\label{subsec:haac_vs}

\begin{figure}[t!]
\centerline{\includegraphics[width=.97\columnwidth]{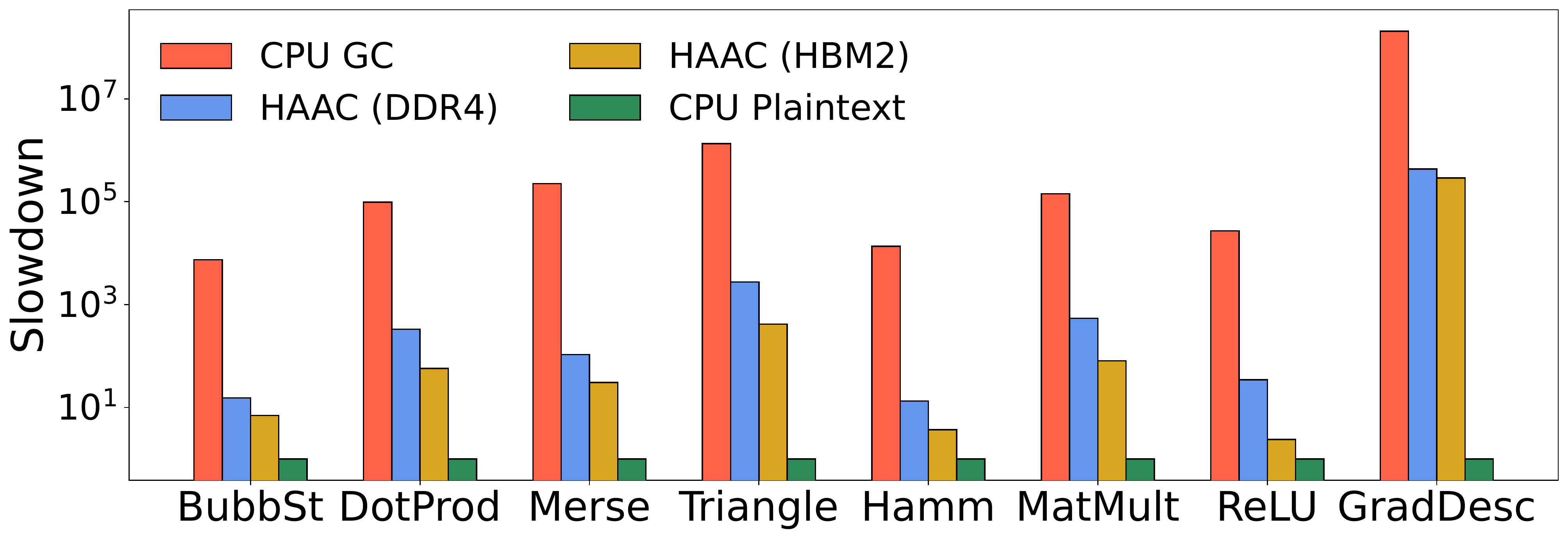}}
\vspace{-1em}
\caption{
GC slowdown normalized to CPU plaintext (shown as 1).
Comparisons include EMP on the CPU (CPU GC) and 
a 16 GEs, 2MB SWW HAAC accelerator (DDR4 and HBM2) under optimal reordering.}
\vspace{-1.5em}
\label{fig:cpu_haac_plain}
\end{figure}

Understanding speedup relative to a software implementation helps show the merit of an accelerator.
However, in the end, the ultimate measure of performance in PPC is how well it performs relative to non-encrypted plaintext.
Figure~\ref{fig:cpu_haac_plain} compares the runtimes
of HAAC (using both DDR4 and HBM2) 
to native, plaintext C++ and EMP (see CPU GC).
There will always be a cost associated with secure computing.
With increased data sizes and work-per-function, there is simply more work to be done, and the same holds true for other PPCs.
Understanding the cost is important in deciding when and where to deploy secure computing.
We further argue that the results are encouraging.

Compared to CPU-run GCs, a HAAC accelerator with DDR4 achieves a geomean speedup of 589$\times$.
We report this number at the beginning of our paper as it assumes the same off-chip bandwidth as the benchmarked CPU.
A HAAC accelerator with HBM2 has a geomean speedup of 2,627$\times$ 
assuming the best reordering scheme for each benchmark,
while the geomean slowdown compared to plaintext is 76$\times$.
GradDesc is particularly slow.
This is because performing floating point securely with GCs is very expensive (the Boolean circuits are complex) while the plaintext CPU can process floating point quickly using native instructions. 
Considering only integer benchmarks, the geomean slowdown compared to plaintext is only 23$\times$.
Overall, HAAC eliminates most of the performance overheads of GC, 
making the slowdown much more tolerable.
Additional compiler optimizations, higher levels of parallelism (e.g., multiple HAAC cores), and processing-in-memory (PIM) may help close the gap.

\subsection{Comparing to Related Work}

We conclude by comparing HAAC against prior work.
We note that prior work (that uses AES) uses the less secure fixed-key GCs setup~\cite{10.1007/978-3-030-56880-1_28}.
Some use SHA-1 instead of AES, which is simpler and less secure.
Comparisons use results from 
published papers, 
and if prior work was modified to support the more computationally expensive construction, we would expect our results to improve.
We acknowledge that the comparison has challenges.
Beyond the algorithms used, ASICs have higher frequency, 
but FPGAs, which many papers use, have large dies, e.g., 84 mm$^2$~\cite{FPGAarea}.
Moreover, most prior work uses small benchmarks that do not stress off-chip bandwidth,
which is one of HAAC's primary contributions.
We present a comparison to acknowledge prior work and provide readers with an understanding of what has already been done.

Table~\ref{tab:CompareOther} shows HAAC compares favorably to all prior work.
We use full reordering, a 1MB SWW, and 16 GEs to compare.
FASE~\cite{8735500} is a generic FPGA garbled circuits accelerator that uses a deep pipelined architecture.
MAXelerator~\cite{Hussain2018MAXeleratorFA} is a fixed-function accelerator tailored to process MAC operations only.
FPGA Overlay~\cite{10.1145/3020078.3021746} proposes a FPGA accelerator with a cluster of custom logic implementing AND and XOR gates based on SHA-1.
As can be seen, HAAC outperforms each of them
and performance improvement is especially noticeable for large workloads,
such as AES and matrix multiplication. 

As FASE is general, it is closest to HAAC.
There are two key differences in the compiler and memory.
FASE orders gates by fanout hoping to resolve dependence and reduce stalls. 
HAAC optimizes for ILP first, ordering gates by their level in data dependence graph. 
FASE’s compiler optimization yields a maximum improvement of 9\% whereas HAAC’s full-reordering gives 3.1$\times$ speedup on average.
FASE assumes data fits on-chip.
It uses a dual-port memory for wires and requires 2 reads and 1 write per cycle, introducing systematic stalls. 
HAAC leverages GCs input/output patterns and strides wires across SWW banks to reduce conflicts. 
Optimized FASE hardware spends 1.7 cycle/gate for an 8-bit MAC while HAAC is 0.079 cycle/gate (22$\times$ faster).
The differences are more pronounced with larger workloads. 
First, at 1.7 cycle/gate, FASE is not able to fully utilize a single gate pipeline whereas HAAC can effectively use sixteen. 
Second, the real challenge of GCs computing emerges when considering off-chip effects. 
As the FASE architecture and compiler cannot mask the off-chip random wire reads (OoRW) nor optimize for on-chip wire reuse (SWW), the performance difference will be more pronounced and bandwidth bound. 
These are where the major HAAC contributions shine.
FASE uses an FPGA running at 200MHz.
If we normalize frequency to match HAAC,
we are still 11$\times$ faster than FASE. 
For smaller circuits, like Million-8, HAAC's speedup is limited as the benchmark has \emph{only 33} instructions, leaving little room for optimization.

Others have implemented GCs on GPUs~\cite{10.1145/2523649.2523681, GPU_editdistance, GPU2, Pu2011FastplayAPM}. 
One GPU implementation can garble an average of 75 million gates per second~\cite{10.1145/2523649.2523681} while
HAAC can garble 8.7 billion gates per second.
The power and area of the GPU compared to HAAC is also over 10$\times$ and 70$\times$~\cite{GPU_K20_power}, respectively.
These use a less efficient AND implementation
and will not bridge the  performance gap.

\section{Related Work}
\label{sec:related_work}

\begin{table}[]
\caption{A performance comparison of HAAC against prior work. 
}
\vspace{-1em}
\label{tab:CompareOther}
\resizebox{\columnwidth}{!}{
\setlength{\tabcolsep}{1mm}{
\begin{tabular}{ccccc}
\hline
                                 & Benchmark  & Garbling Time (us) & Our HAAC (us) & Speedup  \\ \hline
\multirow{2}{*}{MAXelerator~\cite{Hussain2018MAXeleratorFA}}     & 5x5Matx-8  & 15.0 (8 cores)       & 1.605         & 9.35     \\
                                 & 3x3Matx-16 & 6.48 (14 cores)    & 1.673         & 3.87     \\ \hline
\multirow{6}{*}{FASE~\cite{8735500}}            & AES-128    & 439              & 3.607         & 122      \\
                                 & Mult-32    & 52.5               & 1.246         & 42.1     \\
                                 & Hamm-50    & 3.35              & 0.219         & 15.3     \\
                                 & Million-8  & 1.30              & 0.218         & 5.94     \\
                                 & 5x5Matx-8  & 438            & 1.605         & 273      \\
                                 & 3x3Matx-16 & 378                & 1.673         & 226      \\ \hline
\multirow{4}{*}{FPGA Overlay~\cite{10.1145/3020078.3021746}}    & Add-6      & 2.80                & 0.136         & 20.6     \\
                                 & Mult-32    & 180                & 1.246         & 144      \\
                                 & Hamm-50    & 14.0                 & 0.219         & 63.9     \\
                                 & Million-2  & 0.950               & 0.062         & 15.3     \\ \cline{2-5}
~\cite{8945639}               & 5x5Matx-8  & 9.66E+04              & 1.605         & 6.02E+04 \\ \cline{2-5}
\multirow{4}{*}{~\cite{8916407}} & Add-16     & 253                & 0.396         & 639      \\
                                 & Mult-32    & 2.38E+04              & 1.246         & 1.91E+04 \\
                                 & Hamm-50    & 1.55E+03               & 0.219         & 7.08E+03 \\
                                 & 5x5Matx-8  & 1.84E+05             & 1.605         & 1.15E+05 \\ \hline
GPU~\cite{10.1145/2523649.2523681}                               & AES-128    & 75.0 Gates/us         & 8.70k Gates/us  & 116      \\ \hline
\end{tabular}
}
}
\vspace{-1.5em}
\end{table}

\textbf{GC advancements and implementations:}
Researchers have 
developed new algorithmic techniques to improve GCs efficiency
steadily.
The most used optimizations include 
Point-and-Permute \cite{Beaver90theround}, 
Row Reduction~\cite{10.1145/336992.337028,10.1007/978-3-642-10366-7_15},
FreeXOR~\cite{inproceedingsFreeXOR},
and Half-Gate~\cite{10.1007/978-3-662-46803-6_8_HalfGate}.
Point-and-Permute reduces table rows the Evaluator decrypts but increases table size.
Row reduction reduces the number of table rows and is the predecessor to the Half-Gate optimization.
Several software implementations of GCs now 
exist~\cite{ObliVM, inprocFrigate, inproceedingsGC, emp-toolkit}.
They offer similar utility but differ in interfaces and programming language.

\textbf{GCs accelerators:} Prior work has looked at accelerating GCs with FPGAs~\cite{Hussain2018MAXeleratorFA,Songhori2016GarbledCPUAM,10.1145/3020078.3021746,8916407} as well as GPUs~\cite{Pu2011FastplayAPM,10.1145/2523649.2523681,GPU_editdistance, GPU2}.
To the best of our knowledge HAAC is the first ASIC GCs accelerator. 
Prior work does not consider parallel processing units and pipelining at the same time, and most do not optimize for off-chip communication.
HAAC outperforms all prior accelerator and GPU work, see Section~\ref{sec:evaluation} for details.

\textbf{Other PPC accelerators and hardware designs:} Most work to date has focused on accelerating HE.
HE also incurs extremely high overheads, and recent work has made significant advances in mitigating these overheads for integer schemes~\cite{feldmann2021f1, reagen2020cheetah, bts, craterlake, riazi2020heax, CHOCO-TACO}.
Others have looked at accelerating Boolean HE.
However, Boolean HE evaluation is tens to hundreds of times slower than GCs~\cite{TFHE-boots, TFHE-encoding, MATCHA}.
E.g., computing a single TFHE addition takes 3.51s~\cite{TFHE_cpu_gpu}, while it takes only 12.1us with GCs.
Comparing accelerated solutions, HAAC can evaluate GCs gate with a maximum latency of 18ns while TFHE ASICs take 0.18ms per gate~\cite{MATCHA}.
Some have also looked at running HE kernels on GPUs~\cite{cuFHE, reagen2020cheetah}.
Two orders of magnitude speedup over a CPU are typical,
which is impressive but short of mitigating the 5-6 order of slowdown.

Lastly, we note that the SWW is similar in spirit to Mill processor's Belt~\cite{MillCPU}.

\section{Conclusion}
\label{sec:conclusion}

This paper demonstrates the efficacy of co-design in accelerating garbled circuits.
Our approach, named HAAC, includes a compiler, ISA, and microarchitecture designed together in order to significantly improve GCs performance and efficiency while maintaining generality and programmability.
We show how many of the complexities typical of high-performance hardware can be avoided by striking a balance in responsibilities between the hardware and software, with the compiler handling scheduling, data layout, and data movement.
Our specific contributions are the development of gate engine (GE) to speedup gate computations, unique memory structures tailored to the needs of each GCs data structure, a compiler that produces high-performance mappings of high-level code onto the hardware,
and the SWW to capture wire reuse without tagging logic.
While this paper focuses on GCs, we expect HAAC's VLIW-style co-design philosophy to be generally applicable to other data oblivious workloads, broadening its impact.
\begin{acks}
This work was supported in part by the Applications Driving Architectures (ADA) Research Center, a JUMP Center co-sponsored by SRC and DARPA.
We express our gratitude to Marshall Ball (NYU) for his insights on Garbled Circuits. 
We are also grateful to Mohammed Nabeel (NYU) for his assistance with SRAM and Design Compiler. 
Additionally, we thank Xiao Wang (Northwestern) for his guidance on the EMP-Toolkit. 
Finally, we would like to acknowledge the anonymous reviewers for their thorough and insightful feedback. 
The views, opinions, and/or findings expressed are those of the authors and do not necessarily reflect the views of the sponsors.
\end{acks}


\bibliographystyle{ACM-Reference-Format}
\balance
\bibliography{sample-base}

\end{document}